\documentclass[aps, onecolumn,superscriptaddress,footinbib]{revtex4}

\usepackage[T1]{fontenc}
\usepackage[latin9]{inputenc}
\usepackage[letterpaper]{geometry}
\usepackage{float}
\usepackage{graphicx}
\usepackage{amssymb}
\usepackage{ae} 

\usepackage{morefloats}
\usepackage{graphicx,subfigure,epsfig,epstopdf}
\usepackage{dcolumn}

\usepackage{amsmath,amssymb}
\usepackage{float}
\usepackage{subfigure}
\usepackage{fmtcount}
\usepackage{upgreek}

\input{epsf}
\input{epsfx}

\def\bra#1{\mathinner{\langle{#1}|}}
\def\ket#1{\mathinner{|{#1}\rangle}}

\begin{document}

\title{Dense Wavelength Division Multiplexed Quantum Key Distribution Using Entangled Photons}

\author{J. Mower}
\affiliation{Department of Electrical Engineering, Columbia University, New York, NY 10027 USA}
\author{F.N.C Wong}
\affiliation{Research Laboratory of Electronics, Massachusetts Institute of Technology, Cambridge, Massachusetts 02139}
\author{J. H. Shapiro}
\affiliation{Research Laboratory of Electronics, Massachusetts Institute of Technology, Cambridge, Massachusetts 02139}
\author{D. Englund}
\affiliation{Department of Electrical Engineering, Columbia University, New York, NY 10027 USA}
\affiliation{Department of Applied Physics and Applied Mathematics, Columbia University, New York, NY 10027 USA}
%

\begin{abstract}
Quantum key distribution (QKD) enables two parties to establish a secret key over a potentially hostile channel by exchanging photonic quantum states, relying on the fact that it is impossible for an eavesdropper to tap the quantum channel without disturbing these photons in a way that can be detected \cite{2002.RMP.Gisn}. Here we introduce a large-alphabet QKD protocol that makes optimal use of temporal and spectral correlations of entangled photons, reaching the maximum number of independent basis states (the Schmidt number) and enabling extremely high information content per photon together with an optimal rate of secret key generation. This protocol, which we call `Dense Wavelength Division Multiplexed Quantum Key Distribution' (DWDM-QKD), derives its security by the conjugate nature of the temporal and spectral entanglement of photon pairs generated by spontaneous parametric down conversion.  By using a combination of spectral and temporal bases, we can adjust the protocol to be resource efficient. We show that DWDM-QKD is well suited to approach the optimal key generation rate using present-day sources, detectors, and DWDM optical networks from classical communications, as well as emerging optical interconnect and photonic integrated chip (PIC) systems.
\end{abstract}

\maketitle

There has been growing interest in QKD schemes employing photons in high dimensional Hilbert spaces, resulting in a potentially very large-alphabet size \cite{2002.RMP.Gisn, 2002.PRL.Cerf-Gisin.d-levelQKD}. Different degrees of freedom have been considered, including temporal \cite{PhysRevA.61.062308,PhysRevLett.98.060503} and spatial modes \cite{KwiatSuperdense}. Ultimately, the number of bits per photon is limited by the number of independent basis states spanning the Hilbert space, given by the Schmidt decomposition. While previous large-alphabet protocols in the temporal basis have reached up to 16 time bins \cite{PhysRevLett.98.060503}  with a bit error rate of 5\%, they were not able to reach the Schmidt number because, in practical situations, the detector timing jitter greatly exceeds the correlation time of entangled photons generated by typical spontaneous parametric down conversion (SPDC) sources. We overcome this limitation by employing temporal and spectral correlations simultaneously to match the performance of present-day detectors. Thus, given a certain detected-pair flux $n$ and phase-matching bandwidth $\Delta \Omega$, the DWDM-QKD protocol proposed here allows two parties, Alice and Bob, to generate a secure key at the maximum rate possible in the time-frequency bases. In particular, DWDM-QKD enables Alice and Bob to generate their shared key at a maximum rate of $n \log_2(\Delta\Omega/n)$. Here, $\Delta \Omega/n$ corresponds to the maximum number of independent time-frequency states in time interval of duration $1/n$, and thus the maximum number of bits per photons is given by $\log_2 (\Delta\Omega/n)$. 

We consider time-frequency entangled photon pairs with a correlation time $\sigma_{cor}$ produced by frequency-degenerate SPDC \cite{PhysRevLett.61.2921}, assuming a pump field at frequency $\omega_p$ with coherence time $\sigma_{coh}$. In the weak pumping limit, the down-converted photon pair can be approximated by the state $\ket{\Psi}=\int^\infty_{-\infty}\int^\infty_{-\infty}\psi(t_{A},t_{B})\ket{t_{A},t_{B}}dt_{A}dt_{B}$, where $\ket{t_{A},t_{B}}=\hat{a}_A^{\dagger}(t_{A})\hat{a}_B^{\dagger}(t_{B})\ket{0}$, and $\hat{a}_{A,B}^\dagger(t_j)$ denote the creation operators at time $t_j$ for Alice and Bob, and $\psi(t_A,t_B)$ is the normalized time-domain biphoton wave function. This state can be written equivalently in the spectral domain, $\ket{\Psi_\omega}=\int^\infty_{-\infty}\int^\infty_{-\infty}\psi(\omega_{A},\omega_{B})\ket{\omega_{A},\omega_{B}}d\omega_{A}d\omega_{B}$, where $\psi(\omega_{A},\omega_{B})=\textnormal{FT}_2\{\psi(t_{A},t_{B})\}$ and FT$_2$ denotes the two-dimensional Fourier transform. In these expressions, time and frequency represent conjugate bases that can be employed to generate a secure key \cite{2002.PRL.Cerf-Gisin.d-levelQKD}. Specifically, we first discretize the biphoton state into $n_t$ orthogonal temporal basis states, $\ket{\sigma_{bin}^i}$, of duration $\sigma_{bin}$, given by $\ket{\sigma_{bin,A}^i,\sigma_{bin,B}^i}=\int_{i\sigma_{bin}}^{(i+1)\sigma_{bin}}\int_{i\sigma_{bin}}^{(i+1)\sigma_{bin}}\psi(t_A,t_B)\ket{t_{A},t_{B}}dt_Adt_B$, and alternatively into $n_\omega$ orthogonal spectral basis states, $\ket{\nu^i}$, of bandwidth $\delta\nu$, given by $\ket{\nu_A^i,\nu_B^i}=\int_{i \delta\nu}^{(i+1)\delta\nu}\int_{i\delta\nu}^{(i+1)\delta\nu}\psi(\omega_A,\omega_B)\ket{\omega_{A},\omega_{B}}d\omega_Ad\omega_B$. We assume $\sigma_{bin}>\sigma_{cor}$ and $\delta\nu>1/\sigma_{coh}$. $n_t$ and $n_\omega$ are bounded by the Schmidt number, $K=n_t n_\omega$. 

In the protocol, Alice creates a biphoton pair, keeps one photon for herself, and sends the other photon to Bob. Alice and Bob randomly switch between measurements in the temporal ($n_t=K$, $n_\omega=1$) and spectral ($n_t=1$, $n_\omega=K$) bases at intervals of $T=1/n$. After a certain time, they publicly compare their basis choices and divide their measurements into three categories, in which they (i) both detected photons and measured in the \emph{same} basis set, (ii) both detected photons and measured in \emph{different} basis sets, and (iii) did not both measure a photon. They discard category (iii) measurements. Type (i) measurements should be perfectly correlated, providing Alice and Bob with the raw sifted key. Eve does not know in which basis Alice and Bob are observing, and will therefore measure in the wrong basis set half of the time
. If, for example, Alice and Bob do spectral measurements and Eve does a temporal measurement, she spreads the photon pair in frequency and increases the error probability in the type-(i) measurements, revealing her presence to Alice and Bob. 

\begin{figure*}
\centering\includegraphics[scale=0.43]{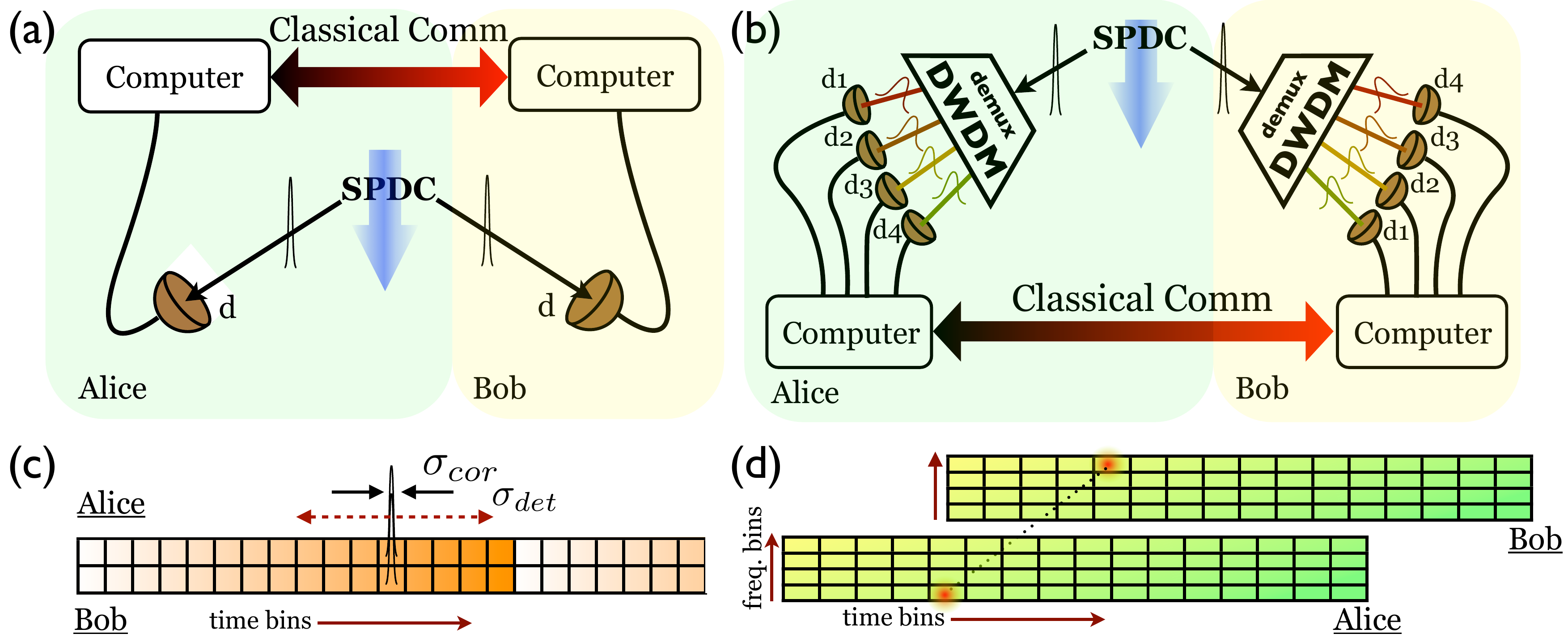}
\centering\caption{(a) The temporal-coding scheme. A strong laser pumps a nonlinear crystal. Photons pairs are generated by SPDC and sent across channels of equal length to Alice and Bob  who measure their arrival times. (b) Time bins agreed upon by Alice and Bob over a public channel. If a photon pair is detected in a given time bin, then that character is shared between Alice and Bob. (c) Alice and Bob place a DWDM before their detectors to obtain spectral information. (d) The new two-dimensional state space.}
\label{schem}
\end{figure*}

The use of two conjugate bases ensures security in the DWDM-QKD protocol. While measurements in these bases is practical for relatively small basis size, the demands on instruments become unrealistic for larger bases; for instance, generating the secure key at 10 bits per photon would require a minimum basis size of 1024, and therefore as many detectors (in a continuously running scheme). The temporal measurements present an additional a challenge because typical SPDC sources at 1550 nm produce photon pairs with $\sigma_{cor} \sim 1$ ps for type-II phase-matched periodically poled KTiOPO$_4$ (PPKTP) ($\sigma_{cor} \sim 0.04$ ps for type-0 phase-matched periodically poled LiNbO$_3$ (PPLN)), which corresponds to a bandwidth of  $\sim$4 nm ($\sim$100 nm). No detector exists to measure such short time bins continuously \footnote{Time-lens approaches \cite{time_lens_lipson2008} may be used to stretch time by some factor $N_{TL}$, but this requires the protocol to be run in bursts so that neighboring intervals do not overlap; up-conversion detectors can offer very short timing resolution, but introduce too many dark counts in the continuously running protocol}. This timing mismatch between practical detectors and practical sources implies that the realizable number of time bins $n_t<K$. When Alice and Bob measure in the temporal domain, they cannot check arrival temporal correlations to their fundamental limit. For instance, if the detector jitter is  $\sigma_{det} \sim 30$ ps, as for SNSPD detectors \cite{4277352}, then $n_t/K \approx 1/30$, as indicated in the temporal measurement in Fig. \ref{schem}a. As a result, Alice and Bob reduce the dimensionality of communication in the temporal domain from the Schmidt number \cite{PhysRevLett.92.127903}, $K\approx\sigma_{coh}/\sigma_{cor}$ to $K'<\sigma_{coh}/\sigma_{det}$, and Eve can obtain finite spectral information without creating observable errors in the temporal basis. This example illustrates that in general, technological limitations reduce the number of basis states far below the fundamental limit.

To solve this problem, Alice and Bob can employ a hybrid basis that constitutes a superposition of temporal and spectral states. They measure in the temporal basis with resolution $\sigma_{bin}$ and simultaneously in the spectral basis with resolution $(\sigma_{bin})^{-1}$. In this case, the maximum number of time bins becomes $n_t=\sigma_{coh}/\sigma_{bin}$ and the number of spectral bins becomes $n_\omega=\sigma_{bin}/\sigma_{cor}$ so that the total alphabet size becomes $n_t n_\omega=\sigma_{coh}/\sigma_{cor}$, which recovers the original Schmidt number, $K$. Now photons are measured in the spectral and temporal basis simultaneously, as shown in Figs. \ref{schem}b and \ref{schem}d. 

The key generation rate depends on the phase-matching bandwidth, $\Delta\Omega$, and the number of photon pairs available for transmission. Alice and Bob can choose to transmit a single pair over this bandwidth, or can split the spectrum into some number of channels, $N_c$, each with photon flux, $n/N_c$. In both cases, the maximum key generation rate evaluates to
\begin{equation}
R \approx n\log_2 \left(\frac{\Delta\Omega}{n}\right).
\label{nomega}
\end{equation}

\begin{figure*}
\begin{centering}
\includegraphics[scale=0.47]{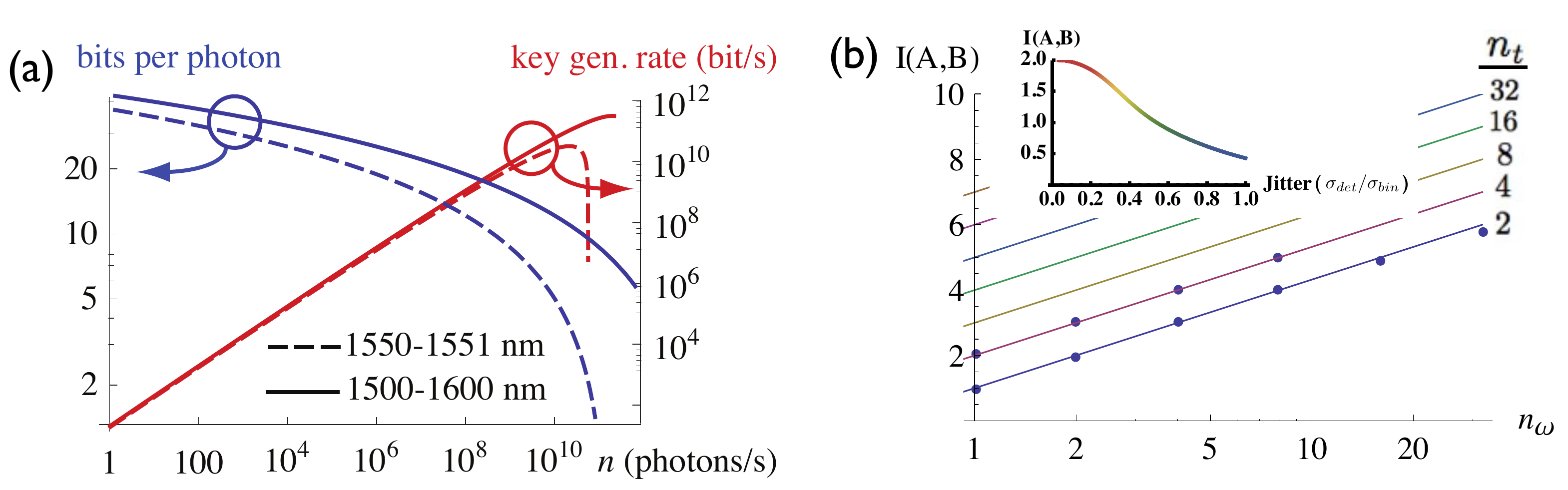}
\par\end{centering}
\caption{\small{(a) Key exchange rate and bits per photon as a function of phase-matching bandwidth and photon flux. (b) Mutual information as a function of the number of spectral channels, $n_\omega$. The individual spectral channels have bandwidth $\delta\nu=\Delta\omega/n_\omega$ and spacing $4\delta\nu$. Increasing the number of spectral channels increases the bits per photon, until the time duration of the filtered photon approaches $\sigma_{bin}$.  The inset shows mutual information as a function of detector timing jitter, $\sigma_{det}$, normalized to $\sigma_{bin}$.}}
\label{jitter}
\end{figure*}

In Fig. \ref{jitter}a, we plot this relationship for two typical sources in the telecom band: a 1-cm type-II phase-matched PPKTP crystal generating photons pairs at 1550 nm with a 4 nm bandwidth; and a 2-cm type-0 phase-matched PPLN crystal at 1550 nm with a 100 nm bandwidth. Alice and Bob can generate a key at almost 20 bpp and 2 Gb/s using the PPLN source.

A careful analysis must take into account the finite overlap of the basis states. When we include this overlap as well as the detector timing jitter, we can calculate the mutual information between Alice and Bob as $I(A,B)=H(A)+H(B)-H(A,B)$, using the Shannon entropy, $H=-\sum_{\{x\}} p^{\{x\}}\log p^{\{x\}}$, where $\{x\}$ is the complete set of indices spanning a probability density function $p$.  We evaluate $I(A,B)$ in Appendix \ref{sec:MI} and plot the results in Fig. \ref{jitter}b, as a function of the number of Gaussian spectral channels, and the number of time bins. We use Gaussian channels to approximate modern DWDM filters. The continuous lines represent the ideal information per photon, and the data points represent our simulated results for the mutual information using a two-dimensional Gaussian,
$\psi(t_A,t_B)\propto e^{-(t_A-t_B)^2/4\sigma_{cor}^2}e^{-(t_A+t_B)^2/4\sigma_{coh}^2}e^{-i\omega_p(t_A+t_B)/2}$,
for the biphoton wave function, where $\omega_p$ is the pump frequency. For a small number of spectral channels, we see good agreement. For large numbers, the Gaussian filters slightly underperform the ideal result due to crosstalk in the closely spaced temporal and spectral bins. This is evident in Fig. \ref{jitter}b for 32 spectral channels with FWHM of $(n_\omega\sigma_{cor})^{-1}$ and channel spacing $4(n_\omega\sigma_{cor})^{-1}$, and two 40 ps time bins. 

Our formalism also enables us to study the effect of detector timing jitter. The results are plotted in the inset in Fig. \ref{jitter}b and show that the mutual information drops rapidly when the timing jitter approaches the time bin duration.

The security check employing conjugate bases in purely frequency and time relies on
the mutually unbiased nature of these bases; a measurement in the wrong basis reveals no information about the state in the other basis and by their conjugate nature introduces errors. We therefore seek a mutually unbiased basis to that employed in our resource-efficient scheme. Alice and Bob's coarse measurements in spectrum are described by  an operator that projects subsets of spectral states onto degenerate eigenvalues $\Omega_l$. The degeneracy within these subsets is lifted by also performing measurements in time, which are described by an operator with coarse timing resolution $T_m$ that correspond to degenerate sets of eigenstates. Simultaneous eigenstates for time and spectral measurements $\ket{\Omega_l T_m}$ describe one basis set. A conjugate basis can be found that forms the second basis for measurements by Alice and Bob. 

\begin{figure}
\centering\includegraphics[scale=0.5]{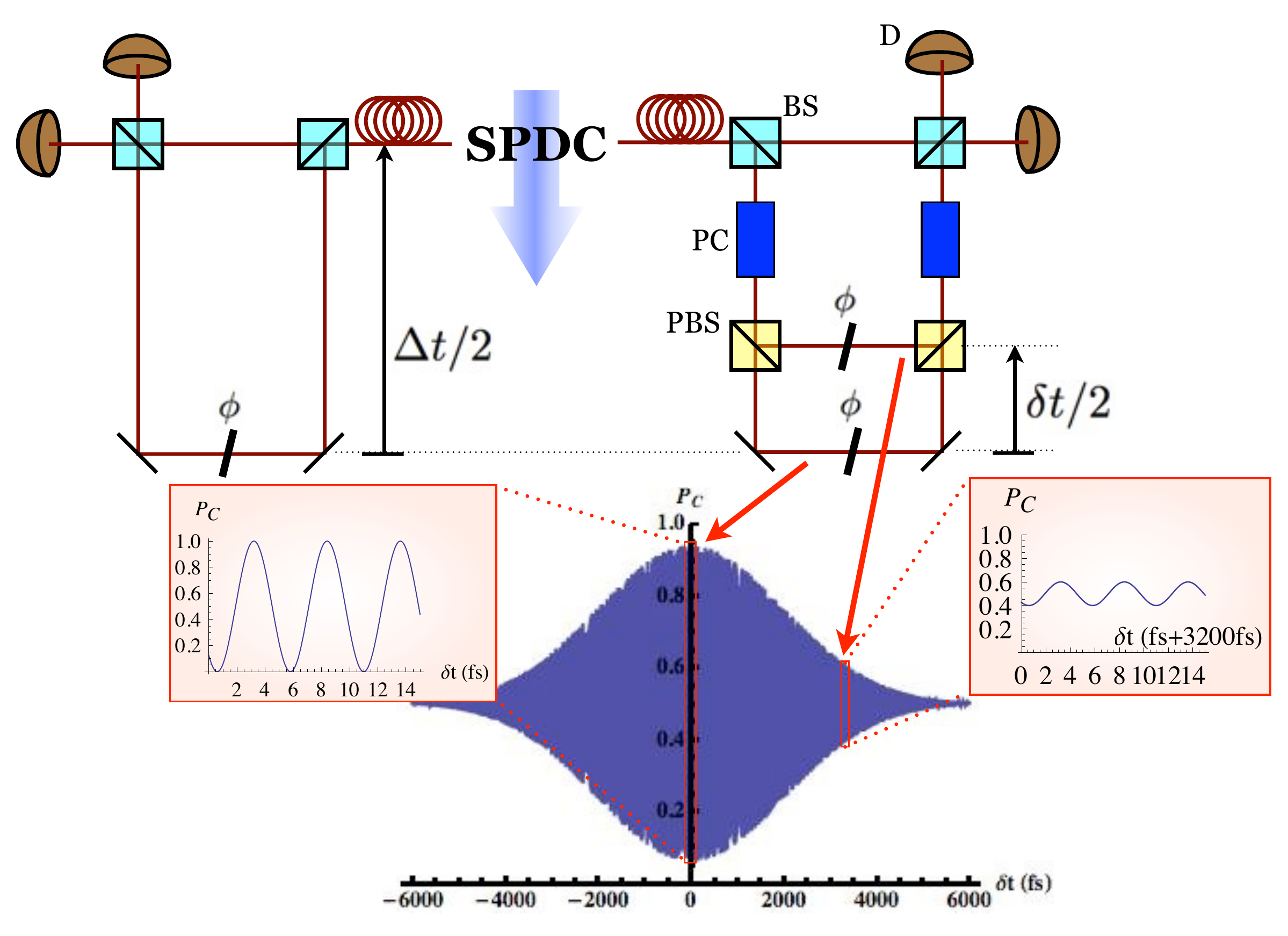}
\centering\caption{\small{The eFI used for security checks. Alice switches the short arm of her Franson between lengths $\delta t_1$ and $\delta t_2$. This allows determination of both $\sigma_{coh}$ and $\sigma_{cor}$ so that weak spectral and temporal measurements on the photon pair can be detected. Alice and Bob do security checks by varying $\delta t$ as shown in the insets. We show one possible switching scheme that makes use of Pockels cells (PC) to rotate the polarization of the photon by $\pi/2$ so that it is either transmitted or reflected at the polarizing beam splitter (PBS) and sent to the extended or standard delay line, respectively. The eFI also includes non-polarizing beam splitters (BS) and single photon detectors (D).}}
\label{secure}
\end{figure}

Under certain conditions, a security check can be performed using simple instrumentation. For now, we assume that Eve chooses to attack by using either a Gaussian envelope in time or one in frequency, i.e.,  
$\hat{E}_t=\int_{-\infty}^\infty e^{-t^2/2(\sigma_{coh}^{E})^2}\ket{t}\bra{t}dt$ \cite{PhysRevLett.98.060503} or $\hat{E}_\omega=\int_{-\infty}^\infty e^{-(\sigma_{cor}^{E})^2(\omega-\omega_p/2)^2}\ket{\omega}\bra{\omega}d\omega$, respectively.
Eve's temporal measurement leads to a decrease in $\sigma_{coh}$, and her frequency measurement creates an increase in the biphoton correlation time, $\sigma_{cor}$. Alice and Bob can detect both of these attacks with the `extended Franson interferometer' (eFI) shown in Fig. \ref{secure}. The eFI is composed of two unbalanced Mach-Zehnder interferometers  (MZI) in the possession of Alice and Bob, where the long path on one arm can be actively modulated.

The probability for Alice and Bob to detect a photon coincidence in their eFI is \cite{PhysRevLett.62.2205}
\begin{equation}
P_C\propto \frac{1}{2}+\frac{1}{2}\cos[\omega(2\Delta t-\delta t)]e^{-\delta t^{2}/8\sigma_{cor}^{2}}e^{-\Delta t^{2}/8\sigma_{coh}^{2}},
\end{equation}
\noindent where $\omega = \omega_p/2$ is the center frequency of the SPDC signal and idler photons, $\Delta t$ is the path-length difference between the long and short arm of Alice's MZI, and $\delta t$ is the path-length difference between Alice's and Bob's long arm. $\Delta t$ is large enough to avoid single photon interference between long and short paths of a single arm of the eFI, and $\delta t$ is varied on the order of $1/\omega$ about zero. The interference is plotted in Fig. \ref{secure} as a function of an additional delay in Alice's MZI. This interference curve shows the oscillations in $P_C$ that is typical of the Franson interferometer near $\delta t = 0$. In addition, this oscillation has a Gaussian envelope whose width is given by $\sigma_{cor}$. 

The visibility of the eFI interference is $V=e^{-\delta t^{2}/8\sigma_{cor}^{2}}e^{-\Delta t^{2}/8\sigma_{coh}^{2}}$. If Eve measures in the temporal domain with a resolution better than $\Delta t$, then Alice and Bob can detect a drop in $V$ near $\delta t =0$; this is the security check used by Kahn \emph{et. al} in Ref.  \cite{PhysRevLett.98.060503}. On the other hand, if Eve measures in the spectral domain with a resolution better than $\Delta\Omega$, then Alice and Bob can detect an increase in $V$ near $\delta t =\sigma_{cor}$. To guard against temporal and spectral measurements by Eve simultaneously, Alice and Bob measure $V$ while Alice switches randomly between delays of 0 and $\sigma_{cor}$ (see Fig. \ref{secure}).

Alice and Bob can deduce the correlation time and coherence time from two visibility measurements $V_1$ and $V_2$ using two delays, $\delta t_1$ and $\delta t_2$, respectively. We label these extrapolated values $\sigma_{coh}^{E'}$ and $\sigma_{cor}^{E'}$, which are given by
\begin{equation}
(\sigma_{cor}^{E'})^2=\frac{1}{8}\frac{\delta t^2_1-\delta t^2_2}{\ln V_2 - \ln V_1}
\end{equation}
\begin{equation}
(\sigma_{coh}^{E'})^2=\frac{1}{8}\frac{\Delta t^2(\delta t^2_1-\delta t^2_2)}{\delta t^2_1\ln V_2 - \delta t^2_2\ln V_1}.
\end{equation}
Using $(\sigma_{coh}^E)^2=1/[(\sigma_{coh}^{E'})^{-2}-\sigma_{coh}^{-2}]$ and $(\sigma_{cor}^E)^2=1/[(\sigma_{cor}^{E'})^{-2}-\sigma_{cor}^{-2}]$ derived from this measurement, the bound on Eve's information per photon is $I_E\le\log_2 (\sigma_{coh}/\sigma_{coh}^E)+\log_2 (\sigma_{cor}^E/\sigma_{cor})$, which is the sum of her information obtained from temporal and spectral measurements. Our assumption of a Gaussian form of Eve's POVM will be generalized in future work. 

\section{Conclusion}
The often limited photon budget for quantum key distribution makes high-dimensional encoding desirable. However, achieving the limit on this dimensionality in the temporal domain using time-frequency entangled photon pairs requires detectors with sub-ps timing jitter and resolution. By invoking conjugate spectral correlations, we present a protocol to approach this fundamental limit using current detectors and existing telecom networks. The conjugate nature of temporal and spectral encoding means that one can trade spectral for temporal bits (and vice versa) to minimize the effect of channel distortion such as nonlinear frequency conversion and dispersion, in addition to optimizing over transmission rate and channel bandwidth. 
\\\\
This work was supported by the DARPA Information in a Photon program, through grant W911NF-10-1-0416 from the Army Research Office.

\section{Methods}
\subsection{Mutual information}
\label{sec:MI}
Alice and Bob ideally communicate information by discretizing the wave function into agreed-upon time-bin $\ket{\sigma_{bin}^i}$ and frequency-bin $\ket{\nu^i}$ macrostates by
\begin{equation}
\ket{\bar{\Psi}}=\sum_{i,j,k,l}G^{i,j,k,l}\ket{\sigma_{bin,A}^{i},\sigma_{bin,B}^{j},\nu_A^k,\nu_B^l},
\end{equation}
where 
\begin{equation}
G^{ijkl}=\int_{i\sigma_{bin}}^{(i+1)\sigma_{bin}}\!\int_{j\sigma_{bin}}^{(j+1)\sigma_{bin}}\textnormal{FT}_2\left[\int_{k\delta\nu}^{(k+1)\delta\nu}\!\int_{l\delta\nu}^{(l+1)\delta\nu}\psi(\omega_A,\omega_B)d\omega_Ad\omega_B\right] dt_A dt_b
\end{equation}. 
The probability of Alice and Bob projecting into time bins $\ket{\sigma_{bin,A}^i}$ and $\ket{\sigma_{bin,B}^j}$ and frequency bins $\ket{\nu_A^k}$ and $\ket{\nu_B^l}$ is $p^{i,j,k,l}=|\bra{\sigma_{bin}^i,\sigma_{bin}^j,\nu_A^k,\nu_B^l}\bar{\Psi}\rangle|^2=|G^{i,j,k,l}|^2$. We label the frequency bins so that for $k=l$, the center frequencies of these bins add to the pump frequency. We plot the mutual information in Fig. \ref{jitter}b as a function of the number of spectral channels added. The wave function is a two-dimensional Gaussian. As we increase the number of spectral channels, the mutual information (MI) increases, however the timing correlations eventually start to decrease, as the filtered photons extend into neighboring time bins. Jitter is also a very important to the MI calculation. We include this in the inset to Fig. \ref{jitter}b. 
\subsection{Detector timing jitter}
Detector timing jitter refers to the added uncertainty in the photon detection time of some stimulus, purely a result of detector electronics. Superconducting nanowire single photon detectors and InGaAs APDs both exhibit jitter of roughly  30 to 40 ps \cite{Hadfield_single_photon}. We model timing jitter as a Gaussian projection, $\hat{\sigma}_{det}=\int e^{-t_x^{2}/2\sigma_{det}^{2}}\ket{t}\bra{t+t_x}dt_x$. The jitter profile of a real photodetector is not truly Gaussian and can be quite asymmetric, however (1) this model allows for first-order analysis and (2) certain single photon detectors do have approximately Gaussian timing jitter \cite{4277352}. If we apply $\hat{\sigma}_{det}$ on both Alice and Bob's photons, assuming the two-dimensional Gaussian given earlier, we get 
\begin{equation}
\hat{\sigma}_{det,A}\hat{\sigma}_{det,B}\ket{\Psi} \propto \int^\infty_{-\infty}\int^\infty_{-\infty}\exp\left[\frac{-(t_A+t_B)^2}{4\sigma_{det}^2+16\sigma_{coh}^2}\right]\exp\left[\frac{-(t_A-t_B)^2}{4\sigma_{det}^2+4\sigma_{cor}^2}\right]e^{i\omega_p(t_A+t_B)/2}\ket{t_{A},t_{B}}dt_{A}dt_{B}
\end{equation}
Since $\sigma_{coh} \gg \sigma_{det}$, the most important effect of jitter is to increase the observed correlation time roughly from $\sigma_{cor}$ to $\sigma_{det}$.  This can have a significant effect on the mutual information between Alice and Bob if $\sigma_{det}$ is on the order of $\sigma_{bin}$, as shown in Fig. \ref{jitter}b. 

\section{Supplementary Information}
\subsection{Lossy Franson interferometry}
\begin{figure}
\begin{centering}
\includegraphics[scale=0.5]{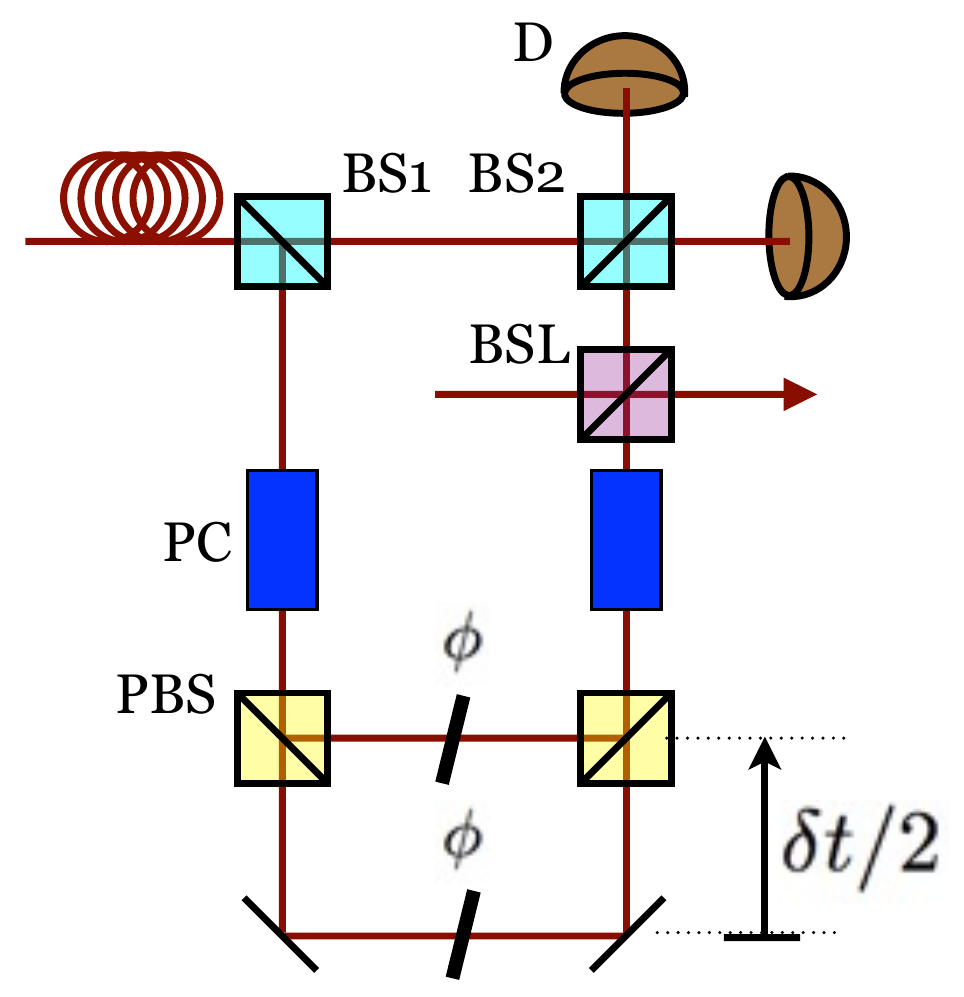}
\par\end{centering}
\caption{\small{The eFI with an additional virtual beam splitter for loss in the long arm.}}
\label{lossy}
\end{figure}

The Franson interference derived in the text assumes lossless propagation through the interferometer. This assumption is not valid in photonic integrated chips or fiber networks. We can account for loss in our analysis by adding a virtual beam splitter in the long path of the otherwise-lossless Franson, which couples the waveguide mode with a vacuum mode (see Fig. \ref{lossy}). We work in the Heisenberg construction, evolving the annihilation operator through the virtual-loss beam splitter and the two Franson beam splitters. The matrix for beam splitters 1 and 2, which leave the third mode undisturbed is given by
\begin{equation}
\hat{U}_i = \left(\begin{matrix} \sqrt{r_i}&\sqrt{1-r_i}&0\\\sqrt{1-r_i}&-\sqrt{r_i}&0\\0&0&1 \end{matrix}\right)
\end{equation}
\noindent where $i\in{1,2}$. The virtual-loss beam splitter is given by
\begin{equation}
\hat{U}_L = \left(\begin{matrix} 1&0&0\\ 0&\sqrt{t_L}&\sqrt{1-t_L}\\0&\sqrt{1-t_L}&-\sqrt{t_L}\\ \end{matrix}\right)
\end{equation}

The resulting annihilation operators are then $\hat{a}_A(t_A)=C_1\hat{a}(t)+C_2\hat{a}(t-\Delta t)$ and $\hat{a}_B(t_B)=C_1\hat{a}(t)+C_2\hat{a}(t-\Delta t - \delta t)$, disregarding the vacuum term, which will not affect coincidence counting. $C_1=\sqrt{r_1}\sqrt{r_2}$ and $C_2=\sqrt{1-r_1} \sqrt{1-r_2} \sqrt{t_L}$. For $r_1=r_2=1/2$, and $t_L=e^{-2t/\tau_\alpha}$ where $\tau_\alpha$ is the lifetime of the photon in the interferometer arm, the visibility simplifies to 
\begin{equation}
V_{PIC}=\frac{2e^{-2\Delta t/\tau_{\alpha}}}{1+e^{-4\Delta t/\tau_{\alpha}}}e^{-\delta t^{2}/2\sigma_{cor}^{2}}e^{-\Delta t^{2}/2\sigma_{coh}^{2}}.
\end{equation}
However for maximum visibility, $C_1=C_2$, so 
\begin{equation}
\frac{\sqrt{r_1}\sqrt{r_2}}{\sqrt{1-r_1} \sqrt{1-r_2}}=\sqrt{t_L}.
\end{equation}
The Franson beam splitters can therefore be tuned to account for loss in the interferometer.

\subsection{Eve and the wave function}
We focus on the case of a single eavesdropper measuring a single photon of the photon pair. Eve's temporal measurement is a Gaussian filtering function
\begin{equation}
\hat{E}_t=\int_{-\infty}^\infty e^{-t^2/2(\sigma_{coh}^{E})^2}\ket{t}\bra{t}dt
\end{equation}
Following \cite{PhysRevLett.98.060503}, the amplitude function 
\begin{equation}
\psi(t_A,t_B)\propto \exp[-(t_A-t_B)^2/4\sigma_{cor}^2]\exp[-t_A^2/4\sigma_{coh}^2]e^{i\omega_p(t_A+t_B)/2}, 
\end{equation}
for $\sigma_{coh}\gg \sigma_{cor}$. 
Therefore 
\begin{eqnarray}
\ket{\Psi_E} &=& \hat{E_t}\ket{\Psi} \\
&\propto& \int^\infty_{-\infty}\int^\infty_{-\infty}\exp\left[-t_A^2\left(\frac{1}{4\sigma_{coh}^2}+\frac{1}{4(\sigma_{coh}^{E})^2}\right)\right]\exp\left[\frac{-(t_A-t_B)^2}{4\sigma_{cor}^2}\right]e^{i\omega_p(t_A+t_B)/2}\ket{t_{A},t_{B}} \nonumber
\end{eqnarray}
so the coherence time of the biphoton packet is strongly influenced by Eve's timing resolution when $\sigma_{coh}^E\ll\sigma_{coh}$. Similarly, we define a weak spectral POVM,
\begin{equation}
\hat{E}_\omega=\int_{-\infty}^\infty e^{-(\sigma_{cor}^{E})^2(\omega-\omega_p/2)^2}\ket{\omega}\bra{\omega}d\omega
\end{equation}
For $1/\sigma_{cor}\gg1/\sigma_{coh}$, $\ket{\Psi}$ can be written in the spectral-domain representation as follows
\begin{equation}
\ket{\Psi}\propto\int\int\exp[-\sigma_{cor}^{2}/4(2\omega_{A}-\omega_{p})^{2}]\exp[-\sigma_{coh}^{2}(\omega_{A}+\omega_{B}-\omega_{p})^{2}]\ket{\omega_{A},\omega_{B}} d\omega_{A}d\omega_{B},
\end{equation}
from which we find that
\begin{eqnarray}
\hat{E}_\omega\ket{\Psi}\propto\int\int\exp[-(\sigma_{cor}^{2}/4+(\sigma_{cor}^{E})^{2}/4)(2\omega_{A}-\omega_{p})^{2}]\\\nonumber
\times\exp[-\sigma_{coh}^{2}(\omega_{A}+\omega_{B}-\omega_{p})^{2}]\ket{\omega_{A},\omega_{B}} d\omega_{A}d\omega_{B}.
\end{eqnarray}
Thus, Eve projects the biphoton pair onto a narrower frequency distribution. 
Reverting to the time-domain representation we get
\begin{equation}
\hat{E}_\omega\ket{\Psi}\propto \int\int\exp(-t_A^2/4\sigma^2_{coh})\exp[-(t_A-t_B)^2/4(\sigma_{cor}^E)^2]e^{i\omega_p(t_A+t_B)/2}\ket{t_A,t_B}dt_Adt_B,
\end{equation}
for $\sigma_{cor} \ll \sigma^E_{cor} \ll \sigma_{coh}$. 


\bibliographystyle{apsrev_no_links.bst}

\end{document}